\documentclass{jpsj-suppl}
\usepackage{txfonts} 
\usepackage{bm}

\title{Inhomogeneity Effects in Topological Superconductors}

\author{Yuki \textsc{Nagai}$^{1}$, Hiroki \textsc{Nakamura}$^{1}$ and Masahiko \textsc{Machida}$^{1}$}

\inst{$^{1}$CCSE, Japan Atomic Energy Agency, 5-1-5 Kashiwanoha, Kashiwa, Chiba, 277-8587, Japan}

\recdate{October 1, 2013}

\abst{
We have constructed a quasiclassical framework on superconductors with strong spin-orbit couplings, applicable to 
Cu$_{x}$Bi$_{2}$Se$_{3}$[Y. Nagai, H. Nakamura, and M. Machida: arXiv:1305.3025].  
The notable point is that in this framework the Bogoliubov-de Gennes Hamiltonians with 
suggested odd-parity pairing states turn to quasiclassical ones with usual spin-triplet Cooper pairs. 
Using this quasiclassical theory, we can investigate inhomogeneity effects such as the phenomena with vortices and surfaces in this 
superconductors and shed light on the pairing state of topological superconductors. 
In this paper, 
we apply the quasiclassical framework to 
the surface bound states with the Dirac-cone energy dispersion originated from the topological invariant in the parent compound Bi$_{2}$Se$_{3}$  in order to 
investigate the robustness of these bound states under the superconducting order parameter. 
The odd-parity gap functions can not open on the Dirac-cone-dispersion band in the Cu-doped Bi$_{2}$Se$_{3}$ 
superconductor. 
We show that the massless Dirac quasiparticles originated from the normal-state topological invariant and the Majorana 
quasiparticles coexist with each other on the surface in the odd-parity topological superconductivity.  
Inhomogeneity effects can be easily investigated with the use of our quasiclassical framework in topological superconductors. 
}

\kword{Topological superconductors, Dirac quasiparticles, Quasiclassical framework}

\newcommand{\imu}{i}

\def\Vec#1{\bm{#1}}
\def\sla#1{\rlap/#1}

\begin{document}
\maketitle

\section{Introduction}

The discovery of topological superconductors has attracted much attention because of new topologically non-trivial states 
of condensed matters. 
Experimentalists have intensively explored evidence of the topological superconductivity by various tools, 
and theorists have debated theoretical framework to describe their various non-trivial superconducting properties\cite{Bernevig15122006,PhysRevLett.105.266401,PhysRevB.76.045302,PhysRevLett.98.106803,RevModPhys.82.3045,PhysRevLett.95.146802,Konig02112007,PhysRevLett.105.14 6801,PhysRevB.75.121306,PhysRevB.81.041309,PhysRevLett.105.136802}. 
Recently, we have constructed a convenient quasiclassical framework for the topological superconductivity characterized by strong spin-orbit coupling and clarified its theoretical correspondence to the spin-triplet superconductivity without the spin-orbit coupling\cite{NagaiQuasi}. 

\begin{figure}[thb]
\begin{center}
\includegraphics[width = 1\columnwidth]{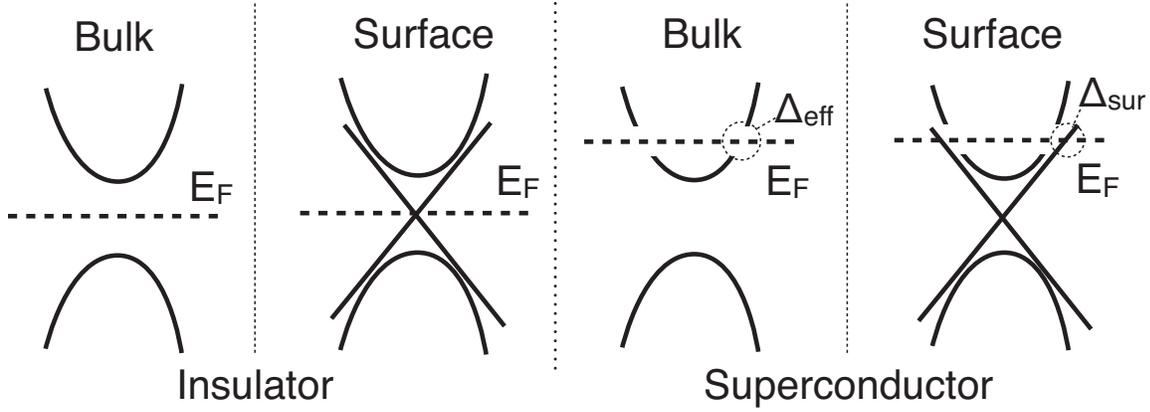}
\caption{\label{fig:fig}
Schematic diagram of the dispersion relations of the surface bound states in the topological insulator and the topological superconductor.
$\Delta_{\rm eff}$ denotes the superconducting gap on the bulk band and $\Delta_{\rm sur}$ denotes that on the surface band. }
\end{center}
\end{figure}

The topological insulator Bi$_{2}$Se$_{3}$ becomes a superconductor with the Cu-doping, and 
Cu$_{x}$Bi$_{2}$Se$_{3}$ has been regarded as a key compound for the investigation of non-trivial topological superconductivity\cite{PhysRevLett.107.217001,PhysRevB.86.064517,PhysRevLett.105.097001,PhysRevB.83.134516,NagaiThermal,NagaiMajo}.
According to the result of the angular photoemission experiment\cite{PhysRevB.86.064517}, 
there are surface bound states originated from the normal-state topological invariant even in the doping material as shown in Fig.~\ref{fig:fig}. 
In this paper, 
we apply the quasiclassical framework to 
the surface bound states with the Dirac-cone energy dispersion originated from the topological invariant in the parent compound Bi$_{2}$Se$_{3}$  in order to 
investigate the robustness of these bound states under the superconducting order parameter. 
With the use of our surface quasiclassical theory, the $8 \times 8$ matrix Dirac Bogoliubov-de Gennes (BdG) equations in the three-dimensional space become 
$2 \times 2$ matrix ones in the two-dimensional space.

\section{Dirac-type Hamiltonian}
Now, let us begin with the massive Dirac type BdG Hamiltonian on
the topological superconductivity expressed as\cite{NagaiThermal,NagaiMajo} 
\begin{equation}
H = \int d{\bm r}
\left(\begin{array}{cc}\bar{\psi}({\bm r}) & \bar{\psi}_{\rm c}({\bm r})\end{array}\right)
\left(\begin{array}{cc}\hat{H}^{-}({\bm r})  & \Delta^{-}({\bm r}) \\
\Delta^{+}({\bm r})& \hat{H}^{+}({\bm r}) \end{array}\right)
\left(\begin{array}{c}\psi({\bm r}) \\
\psi_{\rm c}({\bm r})
\end{array}\right), \label{eq:hami}
\end{equation}
where 
\begin{equation}
\hat{H}^{\pm}({\bm r}) = M_{0}  - \imu  \partial_{x} \gamma^{1} - \imu \partial_{y} \gamma^{2}
- \imu \partial_{z} \gamma^{3} \pm \mu \gamma^{0} .\label{eq:dirac}
\end{equation}
Here, $\gamma^{i}$ is a $4 \times 4$ Dirac gamma matrix,  which can be described as $\gamma^{0} = \hat{\sigma}_{z} \otimes 1$,
$\gamma^{i = 1,2,3} = i \hat{\sigma}_{y} \otimes \hat{s}_{i}$, and $\gamma^{5} = \hat{\sigma}_{x} \otimes 1$ 
with $2 \times 2$ Pauli matrices $\hat{\sigma}_{i}$ in the orbital space and $\hat{s}_{i}$ in the spin space, 
$\psi(\bm r)$ is the Dirac spinor, $\bar{\psi}(\Vec{r}) \equiv \psi^{\dagger}(\Vec{r})\gamma^{0}$, $\bar{\psi}_{c}(\Vec{r}) 
\equiv \psi_{c}^{\dagger} \gamma^{0}$, and $\psi_{c} \equiv {\cal C} \bar{\psi}^{T}$, 
where ${\cal C} (\equiv i \gamma^{2} \gamma^{0})$ is the representative matrix of charge conjugation. 
$\Delta^{-}$ is the gap function and $\Delta^{+} \equiv \gamma^{0} (\Delta^{-})^{\dagger} \gamma^{0}$.
Considering only the on-site pairing interaction, the possible gap form is reduced into six types of functions 
as seen in Table I of Ref.~\cite{NagaiThermal}. 
These gap functions are classified into scalar, pseudo-scalar, and polar vector (four-vector)
associated with the Lorentz transformation,
\begin{equation}
\Delta^{-} = \Delta_{0}, \: \Delta_{0} \gamma^{5}, \: \Delta_{0}\sla{\alpha} \gamma^{5}, 
\end{equation}
where, $\Delta_{0}$ is a scalar magnitude of the gap functions, 
the Feynman slash $\sla{\alpha}$ is defined by $\sum_{\mu} \gamma^{\mu }\alpha_{\mu}$, and 
the gap function including $\sla{\alpha}$ is characterized as a unit four-vector $\alpha_{\mu}$. 
From the Hamiltonian Eq.~(\ref{eq:hami}), the correspondent $8 \times 8$ BdG equations are given as
\begin{equation}
\left(\begin{array}{cc} \hat{h}_{0}({\bm r}) - \mu  & \hat{\Delta}({\bm r})\\
 \hat{\Delta}^{\dagger}({\bm r}) & \hat{h}_{0}({\bm r}) + \mu \end{array}\right)
\left(\begin{array}{c}
u({\bm r}) \\
u_{\rm c}({\bm r})
\end{array}\right)
= E \left(\begin{array}{c}
u({\bm r}) \\
u_{\rm c}({\bm r})
\end{array}\right), \label{eq:bdgdirac}
\end{equation}
where $\gamma^{0} \hat{H}^{\pm} = \hat{h}_{0} \pm \mu$, and $\hat{\Delta} = \gamma_{0} \Delta^{-}$.
Note that $v$ in the conventional Nambu eigen-state form, $(u, v)^T$ is related to $u_{c}$ 
as $v \equiv i \gamma^{2} u_{c}$.  

\section{Surface bound states in normal states}
\subsection{Surface bound states at the $\Gamma$-point }
Topological insulators have gapless quasiparticle states at a surface. 
We consider the surface perpendicular to $z$-axis and the material fills the region of $z > 0$. 
The boundary condition is given by $\Vec{u}(z = 0) = \Vec{u}_{\rm c}(z = 0) = 0$. 
Assuming the translational symmetry along $x$ and $y$, the Dirac Hamiltonian (\ref{eq:bdgdirac}) is expressed as 
\begin{align}
\left[ H_{0}(k_{x},k_{y},- i \partial_{z}) +H_{1}(k_{x},k_{y}) \right] \Vec{u}(k_{x},k_{y}, z) &= \left[ E + \mu \right] \Vec{u}(k_{x},k_{y}, z), \label{eq:diracz}
\end{align}
where $H_{0}(k_{x},k_{y},- i \partial_{z}) \equiv M(k_{x},k_{y},- \imu  \partial_{z}) \gamma^{0} -\imu \partial_{z} \gamma^{0} \gamma^{3}$ and 
$H_{1}(k_{x},k_{y})  \equiv \gamma^{0} (k_{x}  \gamma^{1} + k_{y}  \gamma^{2})$. 
The equation for $\Vec{u}_{c}$ is solved by substituting $\mu \rightarrow - \mu$ into the above equation. 
In the case of $E = -\mu$, the eigenvalue equation with respect to $H_{0}$ becomes 
\begin{align}
\left[M(k_{x},k_{y}, \lambda_{i}) + \sigma \lambda_{i} \right] \Vec{\psi}_{i} &= 0,
\end{align}
%
%
where $\Vec{\psi}_{i}$ is the $i$th eigenvectors of $\gamma^{3}$ with the eigenvalue $\epsilon_{i} = i \sigma$  with $\sigma = \pm 1$ expressed as 
\begin{align}
\epsilon_{1} = i \;, \Vec{\psi}_{1}^{T} &= (1/\sqrt{2})\left(\begin{array}{cccc}0 & i & 0 & 1\end{array}\right), \\
\epsilon_{2} = i \;, \Vec{\psi}_{2}^{T} &= (1/\sqrt{2})\left(\begin{array}{cccc}-i & 0 & 1 & 0\end{array}\right), \\
\epsilon_{3} = -i \;, \Vec{\psi}_{3}^{T} &= (1/\sqrt{2})\left(\begin{array}{cccc}0 & -i & 0 & 1\end{array}\right), \\
\epsilon_{4} = -i \;, \Vec{\psi}_{4}^{T} &= (1/\sqrt{2})\left(\begin{array}{cccc}i & 0 & 1 & 0\end{array}\right).
\end{align}
Here, we assume $\Vec{u}(z) = \sum_{i = 1}^{4} c_{i} \exp [\lambda_{i} z] \Vec{\psi}_{i}$. 
$\lambda_{i}$ dependence of $M(k_{x},k_{y},\lambda_{i})$ determines the existence condition of the bound states, since 
the surface bound states exist only if ${\rm Re}\: \lambda > 0$ under the condition $\lim_{z \rightarrow \infty} \Vec{u}(z) = 0$. 
If $M(\Vec{k}) = M_{0}(k_{x},k_{y}) + M_{1} k_{z}^{2}$, we have $\lambda_{i} = (\sigma \pm \sqrt{1 + 4 M_{0} M_{1}})/(2 M_{1})$ and 
the surface bound state exists when $M_{0} M_{1} < 0$. 
Then, the general solution with respect to $H_{0}$ with $M_{1} < 0$ and $E = -\mu$ becomes 
\begin{align}
 \Vec{u}(k_{x},k_{y}z,E = -\mu) &=  \frac{1}{\sqrt{A}}
\sum_{i = 1}^{2} c_{i} e^{\frac{z}{2 M_{1}}} 
\sinh \left(
K
z \right) 
 \Vec{\psi}_{i}, 
%
\end{align}
where $K = (\sqrt{1 + 4 M_{0} M_{1}})(2 M_{1})$
with $A = \int_{0}^{\infty} dz \exp(z/M_{1}) |\sinh  (K z)|^{2}$.
\subsection{Surface bound states with the rotational symmetry}
We can obtain the eigenvector with respect to $H_{0} + H_{1}$ at $E \neq -\mu$ with the use of the above general solution with respect to $H_{0}$. 
By substituting the above solution  into Eq.~(\ref{eq:diracz}), 
the eigenvalue equations become 
\begin{align}
\left(\begin{array}{cc}
0 & -i k_{+} \\
i k_{-} & 0
\end{array}\right)
\left(\begin{array}{c}
c_{1} \\
c_{2}
\end{array}\right)
&= E' 
\left(\begin{array}{c}
c_{1} \\
c_{2}
\end{array}\right),
\end{align}
with $k_{\pm} \equiv k_{x} \pm i k_{y}$ and $E' \equiv E + \mu$. 
Therefore, we obtain the surface-bound states expressed as 
\begin{align}
\Vec{u}^{\rm N}(k_{x},k_{y},\Vec{r},E') &= \frac{1}{\sqrt{2 A}} e^{i \Vec{k}_{\perp} \cdot \Vec{r}_{\perp}} 
 e^{\frac{z}{2 M_{1}}} 
\sinh \left(
K
z \right) \left[ \Vec{\psi}_{1} + i e^{-i \phi} \: {\rm sgn} (E')\Vec{\psi}_{2} \right], \label{eq:sur}
\end{align}
with $\Vec{k}_{\perp}= (k_{x},k_{y}) = \sqrt{k_{x}^{2}+k_{y}^{2}}(\cos \phi, \sin \phi)$ and $\Vec{r}_{\perp} = (x,y)$.
\subsection{Surface bound states with the six-fold rotational symmetry}
Considering the Hamiltonian for Bi$_{2}$Se$_{3}$ on the triangular lattice\cite{PhysRevLett.107.217001,NagaiMajo}, the eigenvalue equations 
become 
\begin{align}
\left(\begin{array}{cc}
0 & -i P_{+}(k_{x},k_{y}) \\
i P_{-}(k_{x},k_{y}) & 0
\end{array}\right)
\left(\begin{array}{c}
c_{1} \\
c_{2}
\end{array}\right)
&= E' 
\left(\begin{array}{c}
c_{1} \\
c_{2}
\end{array}\right),
\end{align}
with 
\begin{align}
P_{\pm}(k_{x},k_{y}) &\equiv P_{1}(k_{x},k_{y}) \pm i P_{2}(k_{x},k_{y}), \\
P_{1}(k_{x},k_{y})  &\equiv \frac{2}{3} \sqrt{3} \sin \left(\frac{\sqrt{3}}{2} k_{x}\right) \cos \left( \frac{k_{y}}{2} \right), \\
P_{2}(k_{x},k_{y})  &\equiv  \frac{2}{3} \left[
\cos \left(\frac{\sqrt{3}}{2} k_{x}\right) \sin \left( \frac{k_{y}}{2} \right) + \sin \left( k_{y} \right)
 \right].
\end{align}
Therefore, we obtain the surface-bound states on the triangular lattice expressed as 
\begin{align}
\Vec{u}^{\rm N}_{\rm tri}(k_{x},k_{y},\Vec{r},E') &= \frac{1}{\sqrt{2 A}} e^{i \Vec{k}_{\perp} \cdot \Vec{r}_{\perp}} 
 e^{\frac{z}{2 M_{1}}} 
\sinh \left(
K
z \right) \left[ \Vec{\psi}_{1} + i e^{-i \Phi(k_{x},k_{y})} \: {\rm sgn} (E')\Vec{\psi}_{2} \right], \label{eq:surtri}
\end{align}
with 
\begin{align}
e^{-i \Phi(k_{x},k_{y})} &\equiv \frac{P_{1}(k_{x},k_{y}) - i P_{2}(k_{x},k_{y})}
{\sqrt{P_{1}(k_{x},k_{y})^{2} +P_{2}(k_{x},k_{y})^{2}   }}.
\end{align}

%
%


\section{Surface quasiclassical theory}

The quasiclassical theory is founded on an assumption 
that the coherence length $\xi$ is much longer than the  
Fermi wave length $1/k_{\rm F}$ ($\xi k_{\rm F} \gg 1$)\cite{Volovik}.  
The assumption is valid, when the order parameter amplitude $|\Delta_{0}|$ 
is much smaller than the Fermi energy $E_{\rm F}$ ($|\Delta_{0}|/E_{\rm F} \ll 1$), and 
the condition is fully fulfilled in BCS weak-coupling superconductivity.
In this theory, the wave function is expressed by a 
product of the fast oscillating one characterized 
by the Fermi momentum $p_{\rm F}$ and the slowly varying one by the coherence length $\xi$, and  
the quasiclassical solution of the BdG equations is given as 
\begin{align}
\left(\begin{array}{c}
u({\bm r}) \\
u_{\rm c}({\bm r})
\end{array} 
\right) \sim 
\left(\begin{array}{c}
 \Vec{u}^{\rm N}(\Vec{r},\Vec{k}_{\rm F \perp})  f(\Vec{r}_{\perp},\Vec{k}_{\rm F \perp}) \\
 \Vec{u}_{{\rm c}}^{\rm N}(\Vec{r},\Vec{k}_{\rm F \perp})  g(\Vec{r}_{\perp},\Vec{k}_{\rm F \perp})
\end{array}\right),
\end{align}
where $\Vec{u}^{\rm N}$, $\Vec{u}_{{\rm c}}^{\rm N}$ are normal-state eigenvectors at the Fermi level expressed as,  
\begin{align}
 \hat{h}_{0}(\Vec{r}) 
\Vec{u}^{\rm N}(\Vec{r},\Vec{k}_{\rm F \perp}) &= \mu \Vec{u}^{\rm N}(\Vec{r},\Vec{k}_{\rm F \perp}) , \\
\hat{h}_{0}(\Vec{r}) 
\Vec{u}_{{\rm c}}^{\rm N}(\Vec{r},\Vec{k}_{\rm F \perp}) &= -\mu \Vec{u}_{{\rm c}}^{\rm N}(\Vec{r},\Vec{k}_{\rm F \perp})  .
\end{align}
Here, the chemical potential is supposed to be larger than the mass $(\mu > M_{0})$. 
As shown in Ref.~\citen{NagaiQuasi}, there are two solutions in a bulk. 
On the other hand, there is an only one solution at a surface as shown in Eq.~(\ref{eq:sur}).
The eigenvectors are given as 
\begin{align}
\Vec{u}^{\rm N}(\Vec{r},\Vec{k}_{{\rm F} \perp}) &= \Vec{u}^{\rm N}(k_{{\rm F} x},k_{{\rm F} y},\Vec{r},\mu),\\
\Vec{u}^{\rm N}_{\rm c}(\Vec{r},\Vec{k}_{{\rm F} \perp}) &= \Vec{u}^{\rm N}(k_{{\rm F} x},k_{{\rm F} y},\Vec{r},-\mu).
\end{align}
With the use of the above wave function, we reach $2 \times 2$ matrix eigenvalue problem with respect to two functions $(f_{1},g_{1})$ from 
$8 \times 8$ BdG equations. 
The diagonal term is converted as 
\begin{align}
\int_{0}^{\infty} dz \Vec{u}^{\rm N}(\Vec{r},\Vec{k}_{{\rm F} \perp})^{\dagger} (\hat{h}_{0} - \mu) \Vec{u}^{\rm N}(\Vec{r},\Vec{k}_{{\rm F} \perp}) f &= 
- i \Vec{v}_{\rm F \perp} \cdot \Vec{\nabla}_{\perp} f,
\end{align}
with $\Vec{v}_{\rm F} \equiv (\cos \phi, \sin \phi)$ and $\Vec{\nabla}_{\perp} \equiv (\partial_{x}, \partial_{y})$. 
Thus, we have effective two-dimensional $2 \times 2$ quasiclassical BdG equations represented as 
\begin{align}
\left(\begin{array}{cc}
- i \Vec{v}_{\rm F \perp} \cdot \Vec{\nabla}_{\perp}  & \Delta_{\rm sur}(\Vec{r}_{\perp},\Vec{k}_{\rm F \perp}) \\
 \Delta_{\rm sur}(\Vec{r}_{\perp},\Vec{k}_{\rm F \perp})^{\ast} &  i \Vec{v}_{\rm F \perp} \cdot \Vec{\nabla}_{\perp} 
\end{array}\right) 
 \left(\begin{array}{c}
f(\Vec{r}_{\perp},\Vec{k}_{\rm F \perp}) \\
g(\Vec{r}_{\perp},\Vec{k}_{\rm F \perp})
\end{array}\right)
&=  E \left(\begin{array}{c}
f(\Vec{r}_{\perp},\Vec{k}_{\rm F \perp}) \\
g(\Vec{r}_{\perp},\Vec{k}_{\rm F \perp})
\end{array}\right).
\end{align}
All the converted gap functions are listed in Table \ref{table:1}.
As an example exhibited in Table \ref{table:1}, 
the pseudo scalar gap function is equivalent to the spin-triplet 
gap function $\hat{\Delta}_{\rm eff}$ whose $\Vec{d}$-vector rotates in momentum space in the bulk superconductor ($\Vec{d} = (v_{x},v_{y},v_{z})$). Here, $\Vec{v}$ denotes the velocity.  
On the other hand, we should note that the effective surface gap function originated from the pseudo scalar gap function $\Delta_{\rm sur}$ is zero as shown in Table \ref{table:1}, since the off-diagonal term with the pseudo-scalar gap function is converted as 
\begin{align}
\int_{0}^{\infty} dz  \Vec{u}^{\rm N}_{\rm c}(\Vec{r},\Vec{k}_{{\rm F} \perp})^{\dagger} \gamma^{0} \Delta^{-} \Vec{u}^{\rm N}(\Vec{r},\Vec{k}_{{\rm F} \perp}) f &= 
0.
\end{align}
This shows that the normal-state surface bound states written as Eq.~(\ref{eq:sur}) are robust against 
the pseudo-scalar gap functions because the gap can not open as shown in Fig.~(\ref{fig:fig}). 
We should note that even the surface bound states with the six-fold symmetry expressed as Eq.~(\ref{eq:surtri}) are robust, since 
the off-diagonal term with the wave function $\Vec{u}_{\rm tri}^{\rm N}(k_{x},k_{y},\Vec{r},E')$ becomes zero. 
\begin{table*}
\caption{
The correspondence between the original BdG gap functions $\hat{\Delta}^{-}$, the effective ones 
$\hat{\Delta}_{\rm eff}(\Vec{p}_{\rm F})$ and the surface effective ones $\Delta_{\rm sur}(\Vec{k}_{\rm F \perp})$
in quasiclassical theory.
``P-scalar'' denotes a pseudo scalar whose parity is 
odd and ``$i$-polar'' denotes a polar vector pointing the $i$ direction in four dimensional space.}
\label{table:1}
\begin{center}
\begin{tabular}{lcccccl}
\hline
&$\hat{\Delta}^{-}$  &Parity &$\hat{\Delta}_{\rm eff}(\Vec{p}_{\rm F})$  &$\Delta_{\rm sur}(\Vec{k}_{\rm F \perp})$  \\
\hline
Scalar & $\gamma^{5}$ &$+$ &singlet  & 1\\
$t$-polar & $\gamma^{0} \gamma^{5}$ &$+$ & singlet  & 0 \\
P-scalar & $1$ & $-$ & triplet: $\Vec{d} = (v_{x},v_{y},v_{z})$ & 0 \\
$x$-polar & $\gamma^{1}\gamma^{5}$& $-$ &triplet: $\Vec{d} = (0,-v_{z},v_{y})$
& 0
\\
$y$-polar  & $\gamma^{2}\gamma^{5}$ & $-$ &triplet: $\Vec{d} = (v_{z},0,-v_{x})$
& 0
\\
$z$-polar & $\gamma^{3}\gamma^{5}$ & $-$&triplet: $\Vec{d} = (-v_{y},v_{x},0)$& 1 
\\
\hline
\end{tabular}
\end{center}
\end{table*}
\section{Discussion}
\begin{figure}[thb]
\begin{center}
\includegraphics[width = 0.3\columnwidth]{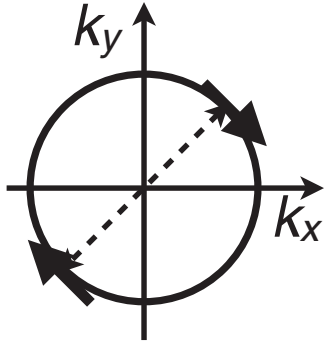}
\caption{\label{fig:kaiten}
Schematic diagram of the spin-momentum locking on the Fermi surface.
}
\end{center}
\end{figure}

We discuss the reason why the surface bound states do not open the superconducting gap due to the 
odd-parity gap functions. 
We note that the spin rotates on the Fermi surface originated from the surface bound states in the normal states expressed as Eq.~(\ref{eq:sur}), which is well known as "spin-momentum locking"\cite{RevModPhys.82.3045}. 
The spin of the quasiparticle with the momentum $\Vec{k}_{\rm F}$ is anti-parallel to that with the momentum $-\Vec{k}_{\rm F}$ 
as shown in Fig.~\ref{fig:kaiten}.
Therefore, the Cooper pairs with parallel spins such as the spin-triplet superconductivity can not form on this 
spin-momentum locking Fermi surface. 
The odd-parity gap functions can not open on the Dirac-cone-dispersion band in the Cu-doped Bi$_{2}$Se$_{3}$ 
superconductor as shown in Fig.~(\ref{fig:fig}). 
This indicates that there are the robust Dirac-type surface states expressed as Eq.~(\ref{eq:sur}) and 
the robust Majorana surface states due to the odd-parity superconductivity. 
Inhomogeneity effects due to the interaction between the Dirac quasiparticles and the Majorana quasiparticles on the 
surface of the topological superconductor 
can be treated by out quasiclassical treatment. 
\section{Conclusion}
We constructed a surface two-dimensional quasiclassical theory which consists of the normal-state surface bound states. 
These surface bound states do not open the superconducting gap due to the odd-parity gap functions, since 
 the Cooper pairs with parallel spins such as the spin-triplet superconductivity can not form on this 
spin-momentum locking Fermi surface. 
We showed that the massless Dirac quasiparticles originated from the normal-state topological invariant and the 
Majorana quasiparticles are coexist on the surface in the odd-parity topological superconductivity. 
Inhomogeneity effects due to the interaction between the Dirac quasiparticles and the Majorana quasiparticles on the 
surface of the topological superconductor 
can be treated by out quasiclassical treatment. 

\section*{Acknowledgment}
The authors would like to acknowledge Yukihiro Ota and for  
helpful discussions and comments. 
This study has been supported by Grants-in-Aid for Scientific Research from MEXT of Japan.

\end{document}